\documentstyle[newarcrc,fleqn,epsf]{article}
\title{Late evolution of  cataclysmic variables\,: \\
the loss of AM Her systems}

\author{F. Meyer, E. Meyer-Hofmeister
\address{Max-Planck-Institut  f\"ur Astrophysik,\\
Karl Schwarzschildstr.~1, D-85740 Garching, Germany}}
        
\begin{document}
\maketitle

\begin{abstract}
The white dwarf in AM Her systems is strongly magnetic and keeps in
synchronous rotation with the orbit by magnetic coupling to the secondary
star. As the latter evolves through mass loss to a cool, degenerate brown dwarf
it can no longer sustain its own magnetic field and coupling is lost. Angular
momentum accreted then spins up the white dwarf and the system no longer
appears as an AM Her system. Possible consequences are run-away
mass transfer and mass ejection from the system. Some of the unusual
cataclysmic variable systems at low orbital periods may be the outcome of this
evolution.

\end{abstract}

\section{Introduction}
In cataclysmic variable systems  the Roche lobe filling secondary star
transfers
mass to the primary white dwarf as the system loses angular momentum by
gravitational wave emission and torques exerted from magnetized winds leaving
the system. For low mass main sequence secondaries the mass - radius exponent
$\zeta$=dlnR/dlnM is about 0.8. Then stellar radius, i.e. Roche
lobe, orbital
separation, and orbital period decrease as mass loss proceeds. If however the
secondary mass has dwindled to about 0.07 $M_\odot$ efficient nuclear energy
production ceases and the secondary becomes a cool partially degenerate brown
dwarf asymptotically reaching the mass-radius exponent -1/3 of an adiabatic gas
sphere. As it passes through the value +1/3 the mean density reaches a
maximum, the orbital period, which is proportional to  its inverse
square root, 
reaches its minimum value and increases again with further mass decrease
[Paczynski and Sienkiewicz 1981, Rappaport et al. 1982, Ritter 1986].

Two problems have appeared here. Firstly, stellar structure calculations yield
a minimum period of 70 minutes while the observed minimum period is close
to 80 minutes. Secondly, since all systems follow the same evolutionary path,
they should accumulate in large number at the minimum period where the  period
becomes stationary in time. This is not observed (for a recent analysis see
Kolb and Baraffe [1999]). 
An assumed increase of the orbital angular momentum loss rate to four times the
gravitational wave value would raise the derived  minimum period to
the observed
value, near 80 minutes \cite{Kolb and Baraffe 1999}. This may point to
residual braking
by the secondary stars magnetic field. Could this magnetic field
differ from star 
to star and thereby spread out the individual period turning points
thus explaining
the missing accumulation of cataclysmic variables (CVs) at one unique minimum
period?

\section{The secondary stars magnetic fields}
A solar type dynamo working between the outer convective envelope and the
radiative interior is thought to cause the high angular momentum loss
and strong
mass transfer for systems at large orbital periods. Its disappearance when
the stars become fully convective is made responsible for the "period gap"
between 3 and 2 hours \cite{Spruit and Ritter 1983, Rappaport et
al. 1982}. But, fully
convective rotating late main sequence stars show chromospheric and coronal
activity indicative of magnetic fields. Their fields have been appealed to for
providing  the friction of dwarf nova  accretion discs in quiescence
\cite{Meyer and
Meyer-Hofmeister 1999a}. A different dynamo working in fully convective stars
may be responsible for these fields. Its long time average, however,
would be the 
same for all secondaries of the same mass and thus also its magnetic braking, 
predicting a unique minimum period.

If however the magnetic fields of the secondaries are confined and held in the 
over-adiabatic structure of these fully convective stars in the same manner as 
suggested for Ap-stars \cite{Meyer 1994}, they are the preserved
relics from an 
earlier dynamo phase and can contain flux to various degrees, as
Ap-stars do.
 
This would allow various degrees of magnetic braking and a spreading of the 
individual period turning points as desired.
Both kinds of magnetic fields, dynamo produced and "fossil", require
convection 
and over-adiabatic structure and these disappear when the stars 
evolve into degenerate adiabatic cool brown dwarfs.

\section{ Past the minimum period: dwarf novae versus AM Her system}

When the magnetic fields of the secondary stars get lost as the systems pass
through their minimum period any residual magnetic braking associated
with it also
disappears. We have suggested \cite{Meyer and Meyer-Hofmeister 1999a}
that the very
low values for the $\alpha$-parameter describing friction in the
quiescent accretion
disks of WZ Sge systems is evidence for this loss of magnetic fields. With
braking reduced to the gravitational wave value the mass transfer rate
is likewise
reduced. This can bring the dwarf novae systems close to and finally beyond the
limit below which they never will burst out again (see the marginal
case of WZ Sge, \cite{
Meyer-Hofmeister et al. 1998}) and make these systems inconspicuous, as has 
often been suggested. But it would possibly not suffice to explain the large
number of "period bouncers" that are expected but not observed
\cite{Patterson 1998}.

How will AM Her systems fare? When the secondaries lose their magnetic field
the white dwarf's magnetic field remains strong at the location of the
secondary.
But all its coupling mechanisms are gone: dipole-dipole interaction, conductive
tying between reconnected field lines of primary and secondary,
convective mixing
in of primary field into secondary envelope [Campbell 1985, 1986,
1989, Lamb 1985,
Lamb and Melia 1988]. Electrical conductivity in the very cool
photosphere of such
brown dwarfs \cite{Allard et al. 1996} becomes poor. Primary fields which
ohmically 
diffuse into the secondary then can not support currents and thus not
transfer coupling stresses between secondary and primary. On the
irradiated side
facing the primary temperatures are higher, conductivity is good but
also prevents
diffusive penetration of the primary's field into the secondary. On
losing its 
coupling the primary white dwarf must spin up driven by the angular momentum
accreted with the transfered mass. 

\section{Spin-up of the white dwarf}

The loss of coupling and the beginning of spin-up has a strong effect on the 
mass transfer rate because the angular momentum that spins up the white 
dwarf reduces the angular momentum of the orbit. This increases the mass 
transfer rate from the  secondary. Then the time scale of mass loss of the 
secondary becomes small (compared to its thermal time scale) and the 
stellar response to mass loss becomes nearly adiabatic, $\zeta$ changes 
from about 1/3 to -1/3, i.e. the star expands as its mass keeps decreasing.
Altogether this significantly raises the mass transfer rate. 

The equations describing this and  the effect of a possible later mass 
loss from the system follow in a standard way by linking polytropic 
structure \cite{Emden 1907}, Roche geometry \cite{Paczynski 1971}, 
mass transfer
through the inner Lagrange point $L_1$ (see Kolb and Ritter [1990] and 
references there), and angular momentum loss by gravitational waves 
(e.g. Misner et al. [1973]). They are

\begin{eqnarray}
\frac{d\dot M}{dt}  = C \dot M^{2/3} \left(\dot M_{\rm{GW}}-X\dot M \right),
\end{eqnarray}
\begin{eqnarray*}
C=
\frac{10^{-21.43}}{k(q)^{1/3}} \left(\frac{M_2}{(0.07M_\odot)}\right)^{-2/3}
P_{80}^{-1/3}~~\rm g^{-2/3}\rm s^{-1/3},
\end{eqnarray*}
\begin{eqnarray*}
\dot M_{\rm{GW}}= \frac{10^{14.79}}{q^{2/3}(1-q)^{1/3}} 
\left(\frac{M_2}{0.07M_\odot}\right)^{8/3} P_{80}^{-8/3}~
~\rm g\,\rm s^{-1},
\end{eqnarray*}
\begin{eqnarray*}
X =  \zeta-\frac{1+\beta q}{3(1+q)}- 2f \sqrt{(1+q)(\frac{r_{\rm LS}}{a}})
   +2(1-\beta q) -(1-\beta)\frac{q}{1+q}\,. 
\end{eqnarray*}

Here $\dot M$ is the mass transfer rate from the secondary, $q$ the mass ratio
$M_2/M_1$, $M_1$ and $M_2$ the masses of primary and secondary star, $k(q)$ a 
factor from the Roche geometry, = 5.97 for $q$ = 0.1 (c.f. Meyer \&
Meyer-Hofmeister  [1983]). $P_{80}$ is the orbital period in units of 80 
minutes. It enters into the polytropic constant by the relation between 
orbital period and mean density \cite{Frank et al. 1985}, for a Roche lobe 
filling secondary and into the gravitational wave term driving the mass
transfer $\dot M_{\rm {GW}}$. $\beta$ is the fraction of the transfered
mass that is 
accreted on the white dwarf, the fraction 1- $\beta$ is lost from the system. 
During spin-up $\beta$=1. The factor $f$ is the specific angular momentum lost 
from the orbit by the matter flow (either accreted on the white dwarf or 
expelled from the system) measured in units of the angular momentum arriving 
with the accretion stream. The latter has been calculated by Lubow \& 
Shu [1975]. $r_{\rm {LS}}$ is the radius of the Kepler orbit around the
white dwarf with this angular momentum and $a$ is the binary
separation.$f$ = 1 for the spin-up phase.

For a positive value of $X$ the solution  of equation (1) tends to the stable 
fixpoint $\dot M$ = $\dot M_{\rm{GW}}/X$. As described above this
value increases
with the onset of spin-up when the response function $X$ becomes smaller with
the change of $\zeta$ and the switch-on of the $f$-term. Had the effective 
driving term $\dot M_{\rm{GW}}$ been four times higher by inclusion of some 
residual magnetic braking before loss of coupling and had dropped to its
value as given above, the change in $X$ would still have increased $\dot M$
by about a factor of three with the onset of spin-up.

We assume that the matter falling towards the white dwarf in the rotating
magnetosphere behaves diamagnetically and experiences a braking force
from the relative motion between field and fluid \cite{King 1993,
Wynn and King 1995}. The strongest interaction occurs at the point
where the free fall path would reach closest approach and the magnetic
field is largest. If the speed of the rotating magnetic field
becomes faster than that of the material there a new phase involving
expulsion of matter can set in.             
								
For our standard case $M_1$ = 0.7$M_\odot$, $M_2$ = 0.07$M_\odot$,
$P_{\rm{orbital}}$ = 
80 minutes, and using the value 0.2 for the square of the radius of 
gyration of the white dwarf (H. Ritter, private communication), the
amount of matter required to reach this 
critical rotation of the white dwarf is $\Delta M$=0.002$M_\odot$,
accumulated in about $6\,10^6$ys.

This idealized case assumes the free fall orbit of Lubow \& Shu's (1975)
calculation. The true orbit will already be affected by the magnetic
interaction and not pass as close to the white dwarf. The critical 
rotation speed will correspondingly be smaller.

Figure 1 shows how the mass transfer rate and the white dwarf spin
develop when magnetic coupling between secondary star and white 
dwarf is lost. A driving term for orbital angular momentum loss of four
times the gravitational wave value was assumed. For comparison the
decrease of mass transfer in dwarf nova systems on loss of residual
magnetic braking is also shown.

\begin{figure}[h]
\epsfxsize=10cm
\centerline{
\epsffile{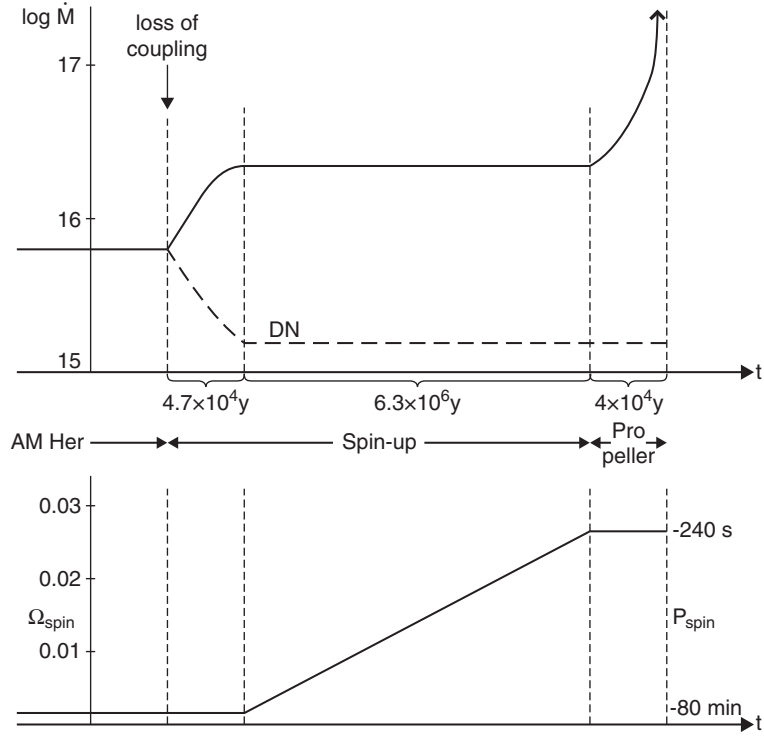}}
\caption{Late evolution of AM Her stars. Upper panel: change of
 mass overflow rate from the companion star $\dot M$ after the loss
 of  magnetic coupling, during the spin-up phase and the propeller
 phase. Lower panel: change of the spin angular frequency of the white
 dwarf}
\end{figure}

\section{The propeller phase}
The falling matter enters the magnetic field of the rotating
dipole. For an inclined dipole the braking and acceleration force
experienced by the matter depends on the rotational phase 
of the dipole. Thus one part of the matter reaches the strong
interaction a bit farther away from the primary and the other comes
closer in. The latter would experience braking and will finally be
accreted. The former can be accelerated outward and flung out of the
system. A propeller effect was already discussed by Illarionov \&
Sunyaev [1975]. The interaction of the magnetic field and the matter
stream is complex. We adapt here results obtained from a model for AE
Aqr by Wynn et al. [1997].

The intermediate polar AE Aqr is observed to spin down at a rate of 
$\dot P_{\rm {spin}}=5.64 \times 10^{-14}$. In the model of Wynn
et al. [1997] the angular momentum given off by the white dwarf adds
to the angular momentum of the stream itself to expell nearly all of
the matter transferred from the secondary. The interpretation of the
observed Doppler tomogram supports this expulsion of matter. Along
these lines we now estimate the orbital angular momentum loss 
$\dot{J}_{\rm{propeller}}$ involved in the expulsion of matter in our
case. We obtain
an estimate for the fraction of matter 1-$\beta$ expelled from the
system by the following consideration. The matter arriving at the
distance of closest approach $r_{\rm{min}}$ needs additional angular
momentum to be accelerated
above escape speed. This is provided by the angular
momentum of the accreted fraction $\beta$.
\begin{eqnarray}
(1-\beta)\left[\sqrt{(1+b^2)} \sqrt{2GM_1 r_{\rm{min}}}
-\sqrt{GM_1r_{\rm{LS}}}\right]= \beta  \sqrt{GM_1r_{\rm{LS}}}
\end{eqnarray}

We take for $r_{\rm{min}}$ the value determined by Lubow \&
Shu [1975] $r_{\rm{min}}= a \cdot \tilde{\omega}_{\rm{min}}$. Assuming that the
matter arrives at infinity with
velocity one half of the escape speed at distance $r_{\rm{min}}$ 
from the white dwarf, $b=
v_{\infty}/v_{\rm {escape}}$= 1/2 one obtains
the estimate
\begin{equation}
\beta=1-\frac {1} {\sqrt{2r_{\rm {min}} /r_{\rm {LS}}}}=0.2.
\end{equation} 
The numerical value results for $q$=0.1.

The orbital angular momentum loss in the propeller phase has to be
taken with respect to the binary's center of gravity.
The forces that produce the roughly $90^\circ$ swing around at the primary
before ejection retard the primary's orbital motion and thereby extract
orbital angular momentum. We use the computed
trajectories from Lubow \& Shu, interpolated for $q$=0.1 and write the
estimate for $\dot{J}_{\rm{propeller}}$
\begin{eqnarray}
\dot{J}_{\rm{propeller}} & = & (1-\beta) \dot M  \left[\sqrt{v_{\rm
{escape}}^2+v_{\infty}^2} ~\cdot r_{\rm {min}}+v_{\infty} \frac{q}{1+q}a
\right],
\end{eqnarray}
where we assume that the expelled matter keeps its angular momentum
with respect to the primary as it rapidly climbs out of the
gravitational potential. The last term on the right side accounts for
the distance between the primary and the binary's center of mass. This yields  
\begin{eqnarray}
\dot{J}_{\rm{propeller}}=\dot M f \sqrt{{GM_1r_{\rm {LS}}}}
\end{eqnarray}
with
\begin{eqnarray}
f=(1-\beta) \sqrt{\frac{2r_{\rm {min}}/a}{r_{\rm {LS}}/a}}
\sqrt{1+b^2} \cdot \left(1+\frac{q}{q+1}\frac{b}{\sqrt{1+b^2}}\frac{1
}{r_{\rm {min}}/a} \right).
\end{eqnarray}

For $q$=0.1 and $\beta$=0.2 one obtains $f$=1.3. A graphical
evaluation of trajectories calculated by Wynn et al. [1997] for AE Aqr
taking into account the smaller mass ratio leads to $f$=1.5.
We emphasize that these estimates are rough. With such values of $f$
the increase of orbital angular momentum loss compared to
$\dot{J}_{\rm {spin-up}}$ changes the sign of $X$ and then leads to
an accelerated
growth of the mass transfer rate [Eq. (1)]. Whether this occurs
depends on the detailed process during the swing around the white dwarf.
For $f$=1.3 and our standard case the time to reach
arbitrarily large $\dot M$ is only $10^{4.6}$ys.
The rate cannot grow indefinitely, finally the magnetic field becomes
unable
to handle the ever growing mass transfer. The
efficiency of acceleration diminishes and angular momentum loss from
the system gets limited.

\section{Further evolution and conclusions}

 What will be the further evolution? If the system stabilizes at a high
mass transfer rate the secondary may lose all its mass in a short time
and a single fast rotating magnetic white dwarf like RE J0317-853
\cite{Barsow et al. 1995} 
will remain. Alternatively, and depending on the field strength of the 
primary, a disc may form which would return the accreted angular 
momentum to the orbit. This would result in a rapid decrease of the
mass transfer rate. But the strong magnetic pressure of the white
dwarf would remove such a disc before the mass transfer rate had
dropped to the stationary value of Eq.(1) possibly resulting in a
cyclic behaviour with phases of high mass transfer rates and rapid 
depletion of the secondary. Perhaps most intriguingly, if expelled 
matter would form a circumbinary disc significant (even unlimited) 
angular momentum could be extracted from the orbit. First 
considerations indicate that nass transfer rates would decrease from 
initially high values to decreasingly lower values on increasingly 
longer time scale as the circumbinary disk evolves. There, even 
some supersoft sources (c.f. van Teeseling \& Kim [1998]) and 
systems like ER Uma with unusally high mass transfer rates 
\cite{Osaki 1996} might find their evolutionary place. But this must be 
left to future investigations. 

This account closely follows a paper to be published in A\&A Letters 
\cite{Meyer and Meyer-Hofmeister 1999b}.

\vspace* {0.3cm}
\it{Acknowledgements}:

\rm{We thank Hans Ritter and Hans-Christoph Thomas for helpful
discussions.
\vspace* {0.3cm}

We thank Brian Warner on his 60th birthday for the continuous
inspiration that his work has brought to cataclysmic variable research.}

\end{document}